\documentclass[11pt, a4paper]{article}

\usepackage{anysize}
\marginsize{3cm}{3cm}{1.5cm}{1.5cm}
\usepackage[T1]{fontenc}
\usepackage{inputenc}

\usepackage{amsfonts}
\usepackage{amssymb}
\usepackage{graphics}
\usepackage{epsfig}
\usepackage{graphicx}
\usepackage{epsfig}
\usepackage{amssymb}
\usepackage{amsfonts}

\usepackage{amsmath}
\usepackage{xcolor}
\usepackage{float}
\usepackage{enumerate}
\usepackage{slashed}
\usepackage{hyperref}
\hypersetup{colorlinks=true,linkcolor=blue,citecolor=blue} 

\numberwithin{equation}{section}

\newcommand {\beq} {\begin{equation}}
\newcommand {\eeq} {\end{equation}}
\newcommand{\bea}{\begin{eqnarray}}
\newcommand{\eea}{\end{eqnarray}}
\newcommand{\bit}{\begin{itemize}}
\newcommand{\eit}{\end{itemize}}
\def\nl{\nonumber \\}

\def\a{\alpha}

\def\b{\beta}

\def\l{\lambda}

\def\p{\partial}

\def\le{\left(}
\def\ri{\right)}

\def\beq{\begin{equation}}
\def\eeq{\end{equation}}

\def\la{{\mathcal{L}}}

\begin{document}

\begin{titlepage}

\begin{flushright}

\end{flushright}
\bigskip
\begin{center}
{\LARGE  {\bf
Volume and complexity for warped AdS black holes
  \\[2mm] } }
\end{center}
\bigskip
\begin{center}
{\large \bf  Roberto  Auzzi$^{a,b}$},
 {\large \bf Stefano Baiguera$^{c}$} {\large \bf and } {\large \bf Giuseppe Nardelli$^{a,d}$} 
\vskip 0.20cm
\end{center}
\vskip 0.20cm 
\begin{center}
$^a${ \it \small  Dipartimento di Matematica e Fisica,  Universit\`a Cattolica
del Sacro Cuore, \\
Via Musei 41, 25121 Brescia, Italy}
\\ \vskip 0.20cm 
$^b${ \it \small{INFN Sezione di Perugia,  Via A. Pascoli, 06123 Perugia, Italy}}
\\ \vskip 0.20cm 
$^c${ \it \small{Universit\`a degli studi di Milano Bicocca and INFN, 
Sezione di Milano - Bicocca, \\ Piazza
della Scienza 3, 20161, Milano, Italy}}
\\ \vskip 0.20cm 
$^d${ \it \small{TIFPA - INFN, c/o Dipartimento di Fisica, Universit\`a di Trento, \\ 38123 Povo (TN), Italy} }
\\ \vskip 0.20cm 
E-mail: roberto.auzzi@unicatt.it, s.baiguera@campus.unimib.it, giuseppe.nardelli@unicatt.it
\end{center}
\vspace{3mm}

\begin{abstract}
We study the Complexity=Volume conjecture  for Warped AdS$_3$ black holes.
We compute the spatial volume of the Einstein-Rosen bridge and we find that its
growth rate is proportional  to the Hawking temperature times the Bekenstein-Hawking entropy.
This is consistent with expectations about  computational complexity
in the boundary theory.
\end{abstract}

\end{titlepage}

\section{Introduction}

 Several quantum information concepts  
have been fruitfully applied to the investigation
of fundamental questions in gravity:
 classical spacetime geometry seems 
somehow to hiddenly encode information properties 
 of a dual quantum system.
Bekenstein-Hawking entropy is indeed proportional 
to the area of the event horizon \cite{Bekenstein:1973ur}
 and the laws of Black Hole (BH) mechanics have a deep connection 
 with thermodynamics  \cite{Bardeen:1973gs}.
The microscopic derivation of the BH entropy in string theory 
given by Strominger and Vafa \cite{Strominger:1996sh},
even  if valid just for some particular extremal cases,
suggests that the BH horizon area should be directly linked
to the number of microstates in some appropriate dual description.

The AdS/CFT correspondence provides  an interesting
theoretical laboratory to investigate quantum information in gravity.
An example which has given us a lot of interesting insights
is the Ryu-Takayanagi construction 
\cite{Ryu:2006bv,Casini:2011kv,Lewkowycz:2013nqa}: 
it links the entanglement entropy 
 of the dual conformal field theory
(CFT) to the geometrical area of a bulk minimal surface 
hanging from the boundary.
Entropy should be somehow related to the counting of 
degrees of freedom in the dual quantum description of a BH.
On the other hand, it turns out that
 entropy is not the correct quantity to focus on
  in order to understand the growth
of the Einstein-Rosen bridge (ERB) in the interior of a BH
\cite{Susskind:2014rva,Susskind:2014moa}. Indeed, the ERB connecting 
two boundaries in an eternal AdS black hole continues 
to grow for a time scale which is much
longer compared to the thermalization scale.

Eternal BHs in AdS space are dual to two entangled copies of the
same CFT living on each of the boundaries \cite{Maldacena:2001kr}.
If we take both the times $t_L, t_R$ of the left and the right CFTs
running forward, this geometry is dual to a time-dependent thermofield doublet state \cite{Hartman:2013qma}.
The size of the ERB connecting the left and the right boundary 
asymptotically grows linearly with time.

A promising candidate to capture this growth
in the dual boundary theory is computational complexity 
\cite{Susskind:2014rva,Susskind:2014moa,Stanford:2014jda,Brown:2015bva,Brown:2015lvg}. 
For a quantum system, it is defined as the minimal number
of elementary unitary operations needed to reach a given quantum
state, starting from an initial reference state.
In the case of quantum mechanics with a finite number of degrees of freedom,
Nielsen and collaborators \cite{Nielsen1,Nielsen2} introduced a nice 
geometrical tool: the problem of quantum complexity is traced back to finding
geodesics in the space of unitary evolutions. In order to extend such analysis 
in quantum field theories, many subtleties arise and 
only recently complexity calculations have been carried out for free
field theories \cite{Jefferson:2017sdb,Chapman:2017rqy,Hashimoto:2017fga}.
Another approach to complexity in quantum field theory 
uses the Liouville action \cite{Caputa:2017yrh} in connection with  
tensor networks \cite{Swingle:2009bg}.

Two different holographic duals of quantum complexity
have been proposed so far: the complexity=volume (CV) conjecture
\cite{Susskind:2014rva,Susskind:2014moa,Stanford:2014jda} and the complexity=action (CA)
conjecture \cite{Brown:2015bva,Brown:2015lvg}.
CV relates complexity to the volume of a codimension one surface
(with maximal volume) anchored at the boundary:
\beq
C_V \propto \, {\rm Max}  \le \frac{V}{G l} \ri \, ,
\eeq
where $G$ is the Newton constant and $l$ the AdS radius.
CA relates complexity to the gravitational action  $I_{WDW}$
evaluated in a  Wheeler-DeWitt patch:
\beq
C_A=  \frac{I_{WDW}}{\pi } \, .
\eeq
Both the conjectures have their own merits.  In particular,
while CV explicitly depends on the AdS curvature $l$, 
CA looks more universal, because no explicit  factors related
to the asymptotic of the space are present.
On the other hand, recent works show that CA seems to overshoot the Lloyd's bound \cite{Lloyd}
at intermediate times  \cite{Carmi:2017jqz} and moreover seems to give 
some curious and weird features: complexity is constant before some initial time $\tau_c$
and immediately after this time $\frac{d C_A}{d \tau}$ is divergent and negative.
On the other hand, CV behaves as a monotonic and smooth function of the time.
Another merit of CV conjecture is that it can be naturally extended to 
consider subregions 
\cite{Alishahiha:2015rta,Ben-Ami:2016qex,Banerjee:2017qti,Abt:2017pmf}.
Another holographic interpretation of the volume was proposed 
in \cite{MIyaji:2015mia}.  

Complexity is interesting not only to capture the linear growth of
ERBs inside BHs;  for example, complexity of formation of BH
 was studied in \cite{Chapman:2016hwi}.
It is also interesting to study complexity 
in connection to other spacetimes; for example, the relation between complexity
and spacetime singularities was investigated in 
\cite{Barbon:2015ria,Bolognesi:2018ion} and
complexity for the AdS soliton was studied in \cite{Reynolds:2017jfs}.
 Complexity for quenches and time-dependent couplings 
was studied in \cite{Moosa:2017yvt,Moosa:2017yiz}.
The effect of dilaton was discussed in \cite{An:2018xhv}.

It is interesting to consider extensions of holography to spacetimes that
are not asymptotically AdS. The most relevant cases for physical
applications would be flat or de Sitter spaces;
unfortunately in these cases very little is known about the 
dual field theory. It is then interesting to study non-trivial modifications
of AdS/CFT where we have more direct information about the structure of the dual.
One of these cases is Warped AdS$_3$/CFT$_2$ \cite{Anninos:2008fx,Detournay:2012pc,Hofman:2014loa,Jensen:2017tnb}, 
which is a correspondence between gravity in  $2+1$ dimensions
in spaces with Warped AdS$_3$ (WAdS) asymptotic and a putative Warped Conformal Field Theory
(WCFT) in $1+1$ dimensions.  In recent times there have been significant progresses in the study of this extension of
the AdS/CFT correspondence. Using warped conformal symmetries, an analog of Cardy formula was derived
in \cite{Detournay:2012pc}; in \cite{Hofman:2014loa} some free examples of WCFT were built.
 Entanglement entropy in WAdS space and in WCFT was studied by several authors, e.g.
\cite{Anninos:2013nja, Castro:2015csg, Azeyanagi:2018har,Song:2016pwx,Song:2016gtd}. 
In this paper we will address the CV conjecture
for BHs in WAdS spaces.

The paper is organized as follows. In Section \ref{BH} we review the Warped black holes solution and
we discuss mass and angular momentum in Einstein gravity; an explicit example
in a theory with ghosts is discussed in appendix \ref{explicit-model}.  In section \ref{CompVol}
we compute the growth rate of the volume of the ERB  as a function of time, both for
the non-rotating and rotating cases. We conclude in  section \ref{Conclu}.

\section{Black Holes in Warped AdS}

\label{BH}

We will be interested in BHs with WAdS$_3$ asymptotic \cite{Moussa:2003fc,Bouchareb:2007yx,Anninos:2008fx}.
This class of metrics should be dual to a boundary WCFT  at finite temperature.
 For the metric we use the notation  of 
\cite{Anninos:2008fx}:
\beq
\label{BHole}
\frac{ds^2}{l^2} = dt^2 +\frac{ dr^2}{(\nu^2+3) (r-r_+)(r-r_-)}
+\le 2 \nu r -\sqrt{r_+ r_- (\nu^2+3)}  \ri dt d \theta +  \frac{r}{4}  \Psi d \theta^2 \, ,
\eeq
\beq
 \Psi(r)= 3 (\nu^2-1)  r +(\nu^2+3) (r_+ + r_-) - 4 \nu \sqrt{ r_+ r_- (\nu^2+3)} \, ,
\eeq
where $0 \leq r  < \infty $, $-\infty < t < \infty$, 
 $ \theta \sim \theta + 2 \pi$ and the horizons are located at $r=r_+, r_-$
with $r_+ \geq r_-$. Taking $ r_+ = r_- = 0 $ we obtain the WAdS spacetime in Poincar\'e patch.
The boundary is parameterized by $(\theta,t)$ and the spatial geometry is flat.
The parameter $\nu$ is related
to the left and right central charges of the boundary WCFT, which in Einstein gravity are \cite{Anninos:2008qb}
\beq
c_L=c_R=\frac{12 l \nu^2 }{G (\nu^2+3)^{3/2}} \, .
\eeq

For $\nu=1$ the  Banados-Teitelboim-Zanelli (BTZ) black hole 
\cite{Banados:1992wn,Banados:1992gq}
is recovered; in this case, the following change of coordinates 
\beq
r=\bar{r}^2 \, , \qquad t = \frac{\sqrt{r_+}-\sqrt{r_-}}{l^2} \bar{t} \, , \qquad 
\theta=\frac{l \bar{\theta} - \bar{t}}{l^2 (\sqrt{r_+}-\sqrt{r_-}) } \, , \qquad r_\pm=\bar{r}_\pm^2 \, ,
\label{change-BTZ}
\eeq
brings the metric to the standard BTZ form:
\beq
ds^2   = - \frac{ \bar{r}^2 - \bar{r}^2_{+} - \bar{r}^2_{-}}{l^2} 
d \bar{t}^2 + \frac{ l^2 \bar{r}^2}{(\bar{r}^2-\bar{r}^2_{+})(\bar{r}^2- \bar{r}^2_{-})} d\bar{r}^2 - 2 \frac{\bar{r}_{+}\bar{r}_{-}}{l} d\bar{t} d\bar{\theta} + \bar{r}^2 d\bar{\theta}^2  \, .
\label{standardBTZ}
\eeq
The case $ \nu=1 $ admits the AdS$_3$ symmetries, i.e.
$ SL(2, \mathbb{R})_L \times SL(2, \mathbb{R})_R$, which
for generic $\nu$ are broken to the WAdS$_3$ symmetry group, that is
$ SL(2, \mathbb{R})_L \times U(1)_R$.

For $\nu^2<1$ the solution is pathological because
it has closed time-like curves. For $\nu^2>1$ the solution is not sick and can
be realized as an exact vacuum
 solution of Topologically Massive Gravity (TMG)
 \cite{Moussa:2003fc,Bouchareb:2007yx},
New Massive Gravity (NMG) \cite{Clement:2009gq}
and also general linear combinations of the two mass terms \cite{Tonni:2010gb}.
We restrict our analysis to the case of positive $\nu$.
So  at the end we will consider just the case  $\nu \geq 1$.

Strictly speaking, the relation between area and entropy
 holds just in Einstein gravity: if we consider higher order
 corrections to the gravitational entropy, we have to 
 use the Wald entropy formula \cite{Wald:1993nt} instead of the geometrical area law.
So the CV conjecture should be
directly applicable just to Einstein gravity and should be
appropriately modified in order to take into account
higher order corrections in the gravitational action.
A proposal for such correction has been put forward in
\cite{Alishahiha:2015rta,Alishahiha:2017hwg}.
The CA conjecture can also be generalized to the case of
higher derivatives corrections to the gravitational action, see
e.g. \cite{Guo:2017rul,Ghodrati:2017roz,Qaemmaqami:2017lzs}.

As far as we know, there is no known non-pathological matter
content in field theory
supporting stretched warped BHs in Einstein gravity \cite{Anninos:2008qb}.
However, they can be obtained as solutions to a perfect
fluid stress tensor with spacelike quadrivelocity \cite{Gurses:1994bjn}. 
Alternatively they can arise as a solution of Chern-Simons-Maxwell electrodynamics 
coupled to Einstein gravity \cite{Banados:2005da,Barnich:2005kq}, 
but a wrong sign for the kinetic Maxwell term is required in order to have solutions with no closed time-like curves 
(which corresponds to $\nu^2 \geq 1$).
Moreover, warped BH can arise in string theory constructions, e.g. 
\cite{Compere:2008cw,Detournay:2012dz,Karndumri:2013dca}.
 In the following we take a pragmatical approach: we suppose that
a consistent  realization of stretched warped BHs in Einstein gravity exists,
and we investigate the CV conjecture.

\subsection{Conserved charges and thermodynamics}

In order to compare with the expectations for complexity,
we need to discuss conserved charges and thermodynamical quantities.
In the Einstein case, the entropy is given by the area of the outer horizon:
\beq
S=S_+=
\frac{l \pi}{4 G} (2 \nu r_+ - \sqrt{r_+ r_- (\nu^2+3)}) \, .
\eeq
At least formally, we can also define the entropy associated to the inner horizon:
\beq
S_-=  \frac{ l \pi}{4 G} 
(  \sqrt{r_+ r_- (\nu^2+3)} - 2 \nu r_- ) \, .
\eeq
The Hawking temperature and the angular velocity are given by \cite{Anninos:2008fx}:
\beq
T= \frac{\nu^2+3}{4 \pi l } \,  \frac{r_+ -r_-}{2 \nu r_+ -\sqrt{(\nu^2+3) r_+ r_-} } \, ,
\qquad
\Omega=\frac{2}{(2 \nu r_+ -\sqrt{(\nu^2+3) r_+ r_-}) l } \, .
\eeq
The first law of thermodynamics gives:
\beq
d M = T dS + \Omega d J \, .
\eeq
Following \cite{Castro:2012av,Giribet:2015lfa}, the existence of a holographic dual implies
 a quantization condition on the product of inner and outer entropies, 
 which in turn must be proportional to the conserved charges of the black hole which are quantized.
Since the angular momentum is the only quantized conserved charge, we obtain $J=S_- S_+ f(\nu)$,
where $f(\nu)$ is a so far arbitrary function which will be fixed by thermodynamics. 

Imposing that the resulting $dM$
is an exact differential, the  function $f(\nu)$ is fixed
and allows to solve for both the conserved charges:
\beq
M=\frac{1}{16 G} (\nu^2+3)
 \le  \le r_{-} + r_{+} \ri - \frac{\sqrt{r_{+}r_{-} (\nu^2 +3)}}{\nu} \ri \, ,
 \label{M guess}
\eeq
\beq
J=\frac{l}{32 G} (\nu^2+3) 
\le \frac{r_- r_+ (3+5 \nu^2)}{2 \nu}
 -(r_+ + r_-) \sqrt{(3+\nu^2) r_+ r_-} 
 \ri \, .
\label{J guess}
\eeq

An explicit realization in Einstein gravity
was discussed in  \cite{Banados:2005da}
and is reviewed in appendix \ref{explicit-model}.
In the formalism of \cite{Banados:2005da,Barnich:2005kq},
 the mass $M$ is identified as
 the conserved quantity associated to the Killing vector
$2 \frac{\p}{\p t}$, while the angular momentum 
is identified as the conserved quantity associated to the Killing
vector $ -\frac{\p}{\p \theta}$.
This provides a non-trivial check that the masses guessed by thermodynamics 
are indeed the same as the ones computed directly in 
an explicit example, which, unfortunately,   has
either closed time-like curves (for $\nu^2<1$) or 
wrong sign Maxwell  term and therefore ghosts (for $\nu^2>1$).

\subsection{Expectations for the asymptotic rate of growth of complexity}


In  \cite{Stanford:2014jda}, it has been proposed that
the asymptotic rate of increase of complexity should be proportional to the product
of temperature times entropy:
\beq
\frac{dC}{d\tau} \simeq T S \, .
\label{stimaTS}
\eeq
The main motivation comes from the fact that complexity growth rate is an extensive
quantity which should have the dimensions of an energy, and which should vanish
for a static object as an extremal BH.
Indeed, for the WAdS BH solutions in Einstein gravity,  we find
\beq
T S = \frac{ (r_+ -r_-)(3+\nu^2)}{16 G} \, .
\eeq
In the next section we will find that the growth rate of the volume of the ERB 
in a WAdS BH is indeed proportional to $TS$.

The authors of \cite{Cai:2016xho} proposed the following bound for the complexity growth rate:
\beq
\frac{dC}{d\tau} \lesssim
  \left[ (M- \Omega J - \Phi Q)_{+} - (M-\Omega J - \Phi Q)_{-} \right] \, ,
\label{bound2}
\eeq
where $ \pm $ indicate that the corresponding values of the quantities are computed at the outer and inner horizons.
With suitable units for complexity, the bound (\ref{bound2}) seems to be saturated in several cases.
For WAdS BHs, the angular velocities computed on the inner and outer horizons are:
\beq
\Omega_+  =\frac{2}{l(2 \nu r_+ -\sqrt{(\nu^2+3) r_+ r_-} )} \, ,
\qquad
\Omega_-  = \frac{2}{l(2 \nu r_- -\sqrt{(\nu^2+3) r_+ r_-} )} \, .
\eeq
If we use the values of mass and angular momentum in eqs. (\ref{M guess})-(\ref{J guess}), we find that
\beq
  (M- \Omega_+ J ) - (M-\Omega_- J )= \frac{ (r_+ -r_-)(3+\nu^2)}{16 G} = TS \, .
\label{TS-MJ}
\eeq
For the purpose of  the case studied in this paper, the saturation of the bound in 
eq.  (\ref{bound2}) is equivalent to eq. (\ref{stimaTS}).

\subsection{Eddington-Finkelstein coordinates}

The Eddington-Finkelstein (EF) coordinates
can be introduced using the light-like geodesics of the metric in eq. (\ref{BHole}).
A system of EF coordinates for the WAdS BH was already introduced in
\cite{Jugeau:2010nq}. The coordinates that we introduce here are not the same,
but  they are still a system of non-singular coordinates at the horizon,
defined using infalling lightlike geodesics,
 that we find convenient for our purposes.
We have the following conserved quantities along geodesics:
\bea
 K &=& 2 \dot{t} + \le 2 \nu r - \sqrt{r_{+}r_{-}(\nu^2 +3)} \ri \dot{\theta} \, ,  \nl
 P &=& \le 2 \nu r - \sqrt{r_{+}r_{-}(\nu^2 +3)} \ri \dot{t} + \frac{r}{2}  \Psi(r) \, ,
\eea
where dots denote derivatives with respect to the geodesic affine parameter.
The null geodesics are found by imposing the additional constraint $ ds^2=0$.
 Solving the equation of motion
  and specializing to $K=0$, we get a particular set of geodesics satisfying
\beq
\dot{t} = \frac{P \le 2 \nu r - \sqrt{r_{+}r_{-}(\nu^2 +3)} \ri  }{(\nu^2 +3)(r-r_{-})(r-r_{+})} \, , \qquad
\dot{\theta} =- \frac{2 P}{(\nu^2 +3)(r-r_{-})(r-r_{+})} \, , \qquad
\dot{r} = \pm P \, .
\eeq
These geodesics can be used to introduce  EF coordinates which are regular at the horizon.

The infalling geodesics correspond to the choice of sign
\beq
\frac{d\theta}{dr} = \frac{2}{(\nu^2 +3)(r-r_{-})(r-r_{+})} \, , \qquad
\frac{d t}{dr} = - \frac{ \le 2 \nu r - \sqrt{r_{+}r_{-}(\nu^2 +3)} \ri }{(\nu^2 +3)(r-r_{-})(r-r_{+})} \, ,
\eeq
and allow to define EF coordinates $(u, \theta_u)$ such that
\beq
du=dt + \frac{  2 \nu r - \sqrt{r_{+}r_{-}(\nu^2 +3)} }{(\nu^2 +3)(r-r_{-})(r-r_{+})}  dr \, , \qquad
d\theta_{u} = d\theta - \frac{2}{(\nu^2 +3)(r-r_{-})(r-r_{+})} dr \, .
\eeq
The finite expression for the coordinate change is
\beq
u= t + r^* (r) \, , \qquad
\theta_{u} = \theta - \frac{2}{(\nu^2 +3)(r_{+}-r_{-})} \log \left| \frac{r -r_{+}}{r-r_{-}} \right| \, ,
\label{TUR}
\eeq
where
\beq
r^* (r) =  \frac{ 2 \nu r_{+} - \sqrt{r_{+}r_{-}(\nu^2 +3)} }{(\nu^2 +3)(r_{+}-r_{-})} \log |r-r_{+}| - \frac{ 2 \nu r_{-} - \sqrt{r_{+}r_{-}(\nu^2 +3)} }{(\nu^2 +3)(r_{+}-r_{-})} \log |r-r_{-}| \, .
\label{RSTAR}
\eeq
In terms of these coordinates, the metric becomes
\beq
\frac{ds^2}{l^2} = du^2 -dr d\theta_{u} + \le 2 \nu r - \sqrt{r_{+}r_{-}(\nu^2 +3)} \ri du d\theta_{u}
+ \frac{r}{4}  \Psi(r)  d \theta_{u}^2 \, .
\label{metricEF}
\eeq

\section{Complexity=Volume }

\label{CompVol}

\subsection{Einstein-Rosen Bridge}

 Kruskal extension for WAdS BHs was studied in \cite{Jugeau:2010nq}.
The Penrose diagrams for WAdS BHs are the same as
the ones for asymptotically flat BHs in $3+1$ dimensional spacetime:
for the special cases $r_-=0$ and $\frac{r_+}{r_-}=\frac{4 \nu^2}{\nu^2+3}$,
the diagram is the same as the one for the Schwarzschild BH,
while for generic $\frac{r_+}{r_-}$ it is  identical to the one for
the  Reissner-Nordstr\"om BH (see figs. 7 and 8 of  \cite{Jugeau:2010nq}).
It is important to emphasize that in the $\nu=1$ case, which is the AdS case, the Penrose diagram 
is different and is the usual AdS one.

As done in \cite{Susskind:2014moa, Stanford:2014jda} for the AdS and the flat cases,
  we consider an extremal codimension one bulk surface
extending between the left and the right side of the Kruskal diagram;
we denote the times at the left and right sides as $t_L, t_R$,  respectively. 
 The dual  thermofield  double  state has the following form:
 \beq
|\Psi_{TFD} \rangle   \propto   \sum_n  e^{-E_n \beta/2- i E_n (t_L + t_R)} | E_{n} \rangle_R  | E_{n} \rangle_L \, ,
\eeq
where $| E_{n} \rangle_{L,R}$ refer to the energy eigenstates of left and right boundary theories,
 $\beta$ is the inverse temperature.
The usual time translation symmetry in Schwarzschild coordinates
corresponds to a forward  time translation on the right side and a backward
translation on the left one \cite{Maldacena:2001kr}, i.e.
\beq
 t_L \rightarrow t_L + \Delta t \, , \qquad t_R \rightarrow t_R - \Delta t \, .
 \eeq
 This corresponds to the invariance of the  thermofield  double state
  under the evolution described by the Hamiltonian 
  $ H=H_L- H_R $ in the associated couple of entangled WCFTs.
If instead we take time running forward on both the copies
of the boundaries, we introduce some genuine time dependence in the problem \cite{Hartman:2013qma}
and the volume of the maximal slice will depend on time \cite{Stanford:2014jda}.
We will then consider the symmetric case with equal boundary times 
\beq
 t_L=t_R=t_b/2 \, .
 \eeq

\subsection{Non-rotating case}

\begin{figure}[ht]  
\begin{center}
\leavevmode
\epsfxsize 5.0 in
\epsffile{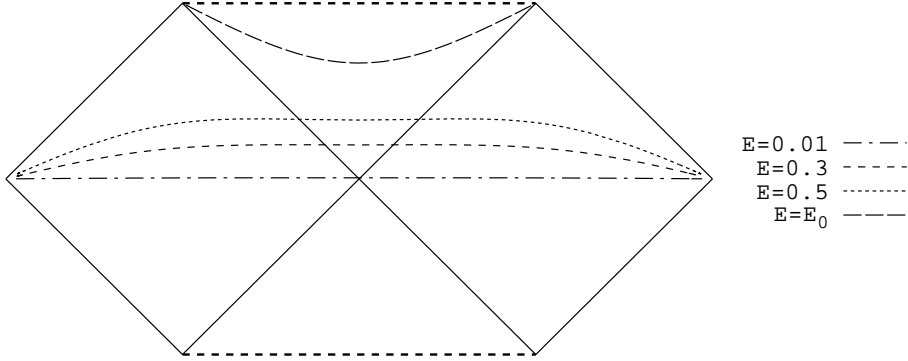}              
\end{center} 
\caption{Solutions to eqs. (\ref{shape}) for the non-rotating case,
 plotted in a Penrose diagram, for $\nu=2.5$ and $r_0=1$. The $E=E_0$ line,
 which sits at constant $r_{\rm min}=  \frac{r_0}{2}$, corresponds to the
 large $t_b$ limit. Penrose diagram coordinates from \cite{Jugeau:2010nq} have been used.}
\label{figura0} 
\end{figure}

 In this section we will compute the volume of the ERB as a function of time \cite{Stanford:2014jda}.
We first study the non-rotating case, setting $r_+=r_0$ and $r_-=0$
in the metric in the coordinates (\ref{metricEF}).
The minimal volume is chosen along the $0 \leq \theta_u \leq 2 \pi$ coordinate, and with
profile functions $u(\l)$, $r(\l)$, written in terms of some parameter $\l$. 
The volume integral will run from some
$\l_{\rm min}$ to some $\l_{\rm max}$, with associated radii $r_{\rm min}$ and $r_{\rm max}$:
\beq
V=2 \cdot 2 \pi \int_{\l_{\rm min}}^{\l_{\rm max}}  d \l \,  l^2 \sqrt{ 
\frac{\dot{u}^2 r}{4} \left[ 3(\nu^2-1) r + (\nu^2+3) r_0 \right]
 - \left(\dot{u} r \nu -\frac{\dot{r}}{2} \right)^2 } = 4 \pi  \int d\lambda \, \mathcal{L} (r,\dot{r},\dot{u})  \, .
\eeq
The factor $2$ takes into account the two sides of the Kruskal extension,
the $2 \pi$ is the result of the integration in $\theta_u$ and the
 dots denote derivatives with respect to $\l$.
The radius $r_{\rm max}$ plays the role of an ultraviolet cutoff;
we will take the limit $r_{\rm max} \rightarrow \infty$
at the end of the calculation.
The conserved quantity from translational invariance in $u$ gives
\beq
E=\frac{1}{l^2}  \frac{\p \la}{\p \dot{u}}=  \frac{\frac{ \nu^2+3 }{4} \dot{u} r (r_0-r)+ \frac{\nu r \dot{r} }{2}}
{\sqrt{ 
\frac{\dot{u}^2 r}{4} \left[ 3(\nu^2-1) r + (\nu^2+3) r_0 \right]
 - \left(\dot{u} r \nu -\frac{\dot{r}}{2} \right)^2 }} \, .
 \label{conserved quantity}
\eeq
It is then useful to gauge the parametrization symmetry  for $\l$ in such a way that 
$V= 4 \pi l^2 \int d \l $:
\beq
\frac{\dot{u}^2 r}{4} \left[ 3(\nu^2-1) r + (\nu^2+3) r_0 \right]
 - \left(\dot{u} r \nu -\frac{\dot{r}}{2} \right)^2  =1 \, ,  \, \qquad 
E=  \frac{ \nu^2+3 }{4} \dot{u} r (r_0-r)+ \frac{\nu r \dot{r} }{2} \, .
\eeq
We can then solve for $\dot{r},\dot{u}$:
\beq
\dot{r}=  2 \sqrt{
\frac{  4 E^2 +  \left(\nu ^2+3\right) r \left(r-r_0\right) }
{ r \left(3 \left(\nu ^2-1\right) r+\left(\nu ^2+3\right) r_0\right)
   } } \, ,
\qquad 
\dot{u}=
\frac{4}{(\nu^2 +3)(r_0-r)} \le \frac{E}{r}  - \frac{\nu}{2}  \dot{r} \ri \, ,
\label{shape}
\eeq
where we took the direction of $\l$ in the direction of increasing $r$.
These equations can be solved numerically; some example of solutions, plotted in a Penrose
diagram, are shown in figure \ref{figura0}.

The minimum radius $r_{\rm min}$ is a solution of $\dot{r}=0$:
\beq
 r_{\rm min}^2- r_0  r_{\rm min} + \frac{4 E^2}{(3+\nu^2)} = 0 \, , \qquad
 r_{\rm min}=\frac{r_0}{2} \le 1 \pm \sqrt{1-\frac{16 E^2}{ r_0^2 (3+ \nu^2) } }  \ri \, ,
 \label{mini}
\eeq
where the physical solution relevant for holographic complexity is the
one with the $+$ sign.  Conventionally,  $t_b=0$ corresponds to $E=0$ and $r_{\rm min}=r_0$.
The $t_b \rightarrow \infty$ limit, instead, corresponds to
coincident roots for $r_{\rm min}$ in eq.~(\ref{mini}), i.e.
 $E\rightarrow  \frac{ r_0}{4} \sqrt{\nu^2 + 3}$
 and $r_{\rm min}=  \frac{r_0}{2} $.
The minimal value of the radial coordinate is inside the black hole horizon 
$\frac{r_0}{2} \leq  r_{\rm min} \leq r_0$.

The volume can be obtained as an integral in $dr$:
\beq 
V = 4 \pi l^2 \int \frac{dr}{\dot{r}}= 2 \pi l^2 \int_{r_{\rm min}}^{r_{\rm max}}  
\sqrt{ 
\frac
{ r \left(3 \left(\nu ^2-1\right) r+\left(\nu ^2+3\right) r_0\right)} 
{  4 E^2+\left(\nu ^2+3\right) r \left(r-r_0\right) }
} dr \, .
\eeq
The difference of $u$ coordinates is:
\bea
\label{DELTAu}
& & u (r_{\rm max}) - u (r_{\rm min}) = \int_{r_{\rm min}}^{r_{\rm max}} dr \frac{\dot{u}}{\dot{r}} 
\nl
&=& \int_{r_{\rm min}}^{r_{\rm max}} dr \left[
\frac{2}{(\nu^2 +3)(r_0-r)} \le \frac{E}{ r }
\sqrt{ 
\frac
{ r \left(3 \left(\nu ^2-1\right) r+\left(\nu ^2+3\right) r_0\right)} 
{  4 E^2+  \left(\nu ^2+3\right) r \left(r-r_0\right) }}
  - \nu  \ri
\right] \, .
\eea
Note that this integral is not divergent for $r \rightarrow r_0$.
The volume can then be written as follows:
\bea
\frac{V}{4 \pi l^2} & = & E( u (r_{\rm max}) - u (r_{\rm min}) )
+ \int_{r_{\rm min}}^{r_{\rm max}} dr
\left\{
\frac{2 \nu E}{(\nu^2+3) (r_0-r)} \right.
\nl & &
\left.
-\frac{\sqrt{r \, 
[4 E^2 -  r (r_0-r)(\nu^2+3)] \,
[(\nu^2+3)r_0 +3 r (\nu^2-1)]} }
{2 (\nu^2+3) r (r_0-r)}
\right\} \, .
\label{volume-u}
\eea
It is important to emphasize that
\beq
\lim_{r_{\rm max}\rightarrow \infty} u(r_{\rm max})- r^* (r_{\rm max})=t_R \, ,
\eeq
is finite  and can be identified with the time at the right boundary.
In the limit $r_{\rm max} \rightarrow \infty$, we can use the explicit expression 
\beq
u(r_{\rm max}) - u (r_{\rm min})  = t_R + r^* (r_{\rm max}) - r^* (r_{\rm min}) \, ,
\label{DELTAu2}
\eeq
obtained specializing eq. (\ref{TUR}) :
\beq
u(r_{\rm max}) =t_R + r^* (r_{\rm max}) \, , \qquad
 u (r_{\rm min})=r^* (r_{\rm min}) \, ,
\eeq
because $t=0$ at $r=r_{\rm min}$ by symmetry considerations.

\begin{figure}[ht]  
\begin{center}
\leavevmode
\epsfxsize 5.0 in
\epsffile{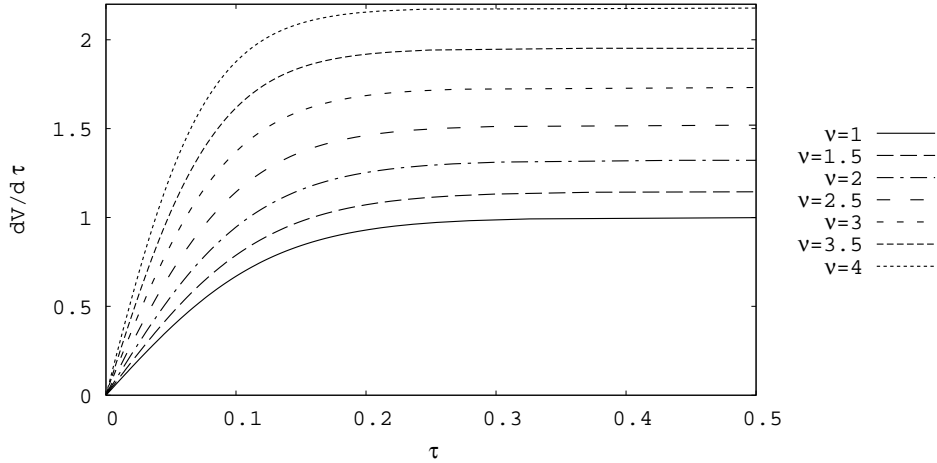}              
\end{center} 
\caption{Time dependence of $\frac{d V}{d \tau}$ in units of $ \pi l$,
for $r_0=1$ and various values of the warping parameter $\nu$.}
\label{figura1} 
\end{figure}

Taking into account that both $E$ and $r_{\rm min}$
depend on $t_R$ (see eq. (\ref{mini}) for the relation among $r_{\rm min}$ and $E$),
the time derivative of eq. (\ref{volume-u}) gives, 
after several cancellations among terms:
\beq
\frac{1}{2 l} \frac{d V}{d t_R}=\frac{d V}{d \tau} = 2 \pi l E \, ,
\eeq
where $\tau=l \, t_b=2 l \, t_R$.
At large $\tau$, $E$  approaches to the constant $E_0=\frac{ r_0}{4} \sqrt{\nu^2 + 3}$.
Computing the constant of motion \emph{E} in eq. (\ref{conserved quantity}) 
for the particular value $ r=r_{\rm min} $ shows that $ E>0$ for $\tau>0$ 
(corresponding to  $\dot{u}>0$) and $ E<0$ for $\tau<0$ (corresponding to $\dot{u}<0$).
 Numerical calculations with the full time dependence can be
 obtained by expressing $\tau$ in terms of $E$ using eqs. (\ref{DELTAu}-\ref{DELTAu2}),
  are shown in figure \ref{figura1}.
 For $\nu=1$ the results in \cite{Stanford:2014jda},
  \cite{Carmi:2017jqz} are recovered, under the change of variables
 in eq. (\ref{change-BTZ}).
 
\subsection{Rotating case}

We use the metric in the coordinates (\ref{metricEF});
the volume functional is:
\beq
V= 4 \pi \int_{\l_{\rm min}}^{\l_{\rm max}} 
 d\lambda \, l^2 \sqrt{\frac{r \dot{u}^2}{4} \Psi
 - \le \frac{\dot{u}}{2} \le 2 \nu r - \sqrt{r_{+}r_{-}(\nu^2 +3)} \ri - \frac{\dot{r}}{2} \ri^2 } = 4 \pi \int d\lambda \, \mathcal{L} (r,\dot{r},\dot{u}) \, .
\eeq
Due to the axial symmetry, the volume is taken along the $\theta_u$ direction.
As before, we find the conserved quantity:
\beq
E = \frac{1}{l^2} \frac{\p \mathcal{L}}{\p \dot{u}} =  \frac{\frac{r \dot{u}}{4} \Psi
-\le \frac{\dot{u}}{2} \le 2 \nu r - \sqrt{r_{+}r_{-}(\nu^2 +3)} \ri - \frac{\dot{r}}{2} \ri \frac12  \le 2 \nu r - \sqrt{r_{+}r_{-}(\nu^2 +3)} \ri
}
{\sqrt{\frac{r \dot{u}^2}{4} \Psi
- \le \frac{\dot{u}}{2} \le 2 \nu r - \sqrt{r_{+}r_{-}(\nu^2 +3)} \ri - \frac{\dot{r}}{2} \ri^2 }} \, .
\eeq
The expression greatly simplifies choosing a parametrization for $ \lambda $ such that
$V= 4\pi l^2 \int d \lambda$, which corresponds to setting
\beq
\frac{r \dot{u}^2}{4} \Psi
- \le \frac{\dot{u}}{2} \le 2 \nu r - \sqrt{r_{+}r_{-}(\nu^2 +3)} \ri - \frac{\dot{r}}{2} \ri^2  = 1 \, .
\label{verde1}
\eeq
This gives:
\beq
E= -  \frac{\nu^2 +3}{4} \dot{u} (r-r_{-})(r-r_{+}) +  \frac{\dot{r}}{4} \le 2 \nu r - \sqrt{r_{+}r_{-}(\nu^2 +3)}  \ri \, .
\label{verde2}
\eeq
Solving eqs. (\ref{verde1}, \ref{verde2}), we obtain the expressions:
\beq
\dot{r} =  2 \sqrt{\frac{4E^2 +  (\nu^2 +3)(r-r_{-})(r-r_{+})}{\le 2 \nu r - \sqrt{r_{+}r_{-}(\nu^2 +3)}  \ri^2 -  (\nu^2 +3)(r-r_{-})(r-r_{+})}} \, ,
\eeq
\bea
& \dot{u} &= \frac{2}{ (\nu^2 +3)(r-r_{-})(r-r_{+})} \nl 
& & \left[ \frac{\sqrt{4E^2 +  (\nu^2 +3)(r-r_{-})(r-r_{+})}
  \le 2 \nu r - \sqrt{r_{+}r_{-}(\nu^2 +3)}  \ri }{ \sqrt{\le 2 \nu r - \sqrt{r_{+}r_{-}(\nu^2 +3)}  \ri^2 -  (\nu^2 +3)(r-r_{-})(r-r_{+})}} 
 - 2 E \right] \, .
\eea
The minimum value $r_{\rm min}$  of the radial coordinate is obtained by solving $\dot{r}=0$:
\beq
r_{\rm min} = \frac{r_{+}+r_{-}}{2} \le 1 \pm \sqrt{1 - \frac{16E^2}{ (\nu^2 +3)(r_{+}+r_{-})^2}} \ri \, .
\eeq
As in the non-rotating case, the physical solution relevant for holographic complexity is the
one with the $+$ sign.  Conventionally, $ t_b=0 $ corresponds to $ E=0 $ and $ r_{\rm min}=r_{+}+r_{-}$. 
The $ t_b \rightarrow \infty $ limit corresponds to $ E \rightarrow  \frac{(r_{+}-r_{-})}{4} \sqrt{\nu^2 +3} $
 and $ r_{\rm min}=\frac{r_{+}+r_{-}}{2}$.

\begin{figure}[ht]  
\begin{center}
\leavevmode
\epsfxsize 5.0 in
\epsffile{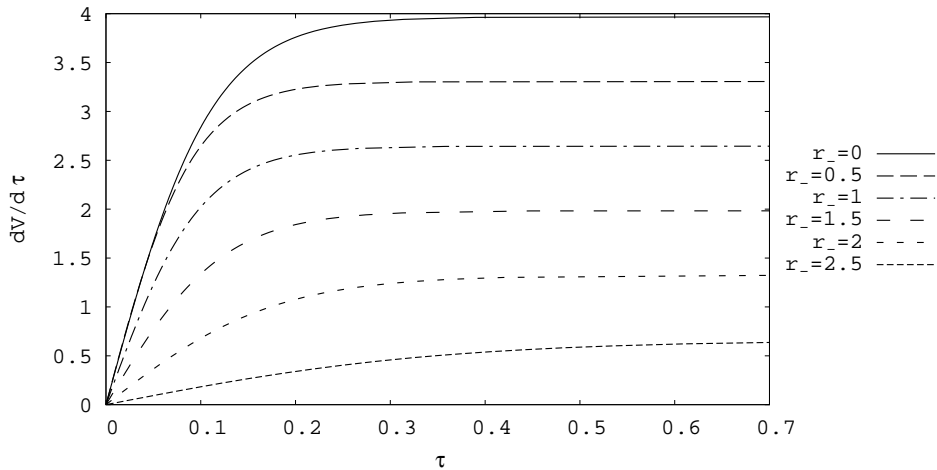}              
\end{center} 
\caption{Time dependence of $\frac{d V}{d \tau}$ in units of $ \pi l$, for $r_+=3$, $\nu=2$ and several values of $r_-$.
For other values of $\nu$ the plots are qualitatively similar.}
\label{figura2} 
\end{figure}

The volume can be expressed as an integral in $dr$ as:
\beq
V= 2 \pi l^2 \int_{r_{\rm min}}^{r_{\rm max}} dr \sqrt{\frac{\le 2 \nu r - \sqrt{r_{+}r_{-}(\nu^2 +3)}  \ri^2 -  (\nu^2 +3)(r-r_{-})(r-r_{+})}{4E^2 + (\nu^2 +3)(r-r_{-})(r-r_{+})}} \, .
\eeq
It is useful to introduce the difference among the extremal values of EF coordinates:
\bea
&& u (r_{\rm max}) - u(r_{\rm min}) = \int_{r_{\rm min}}^{r_{\rm max}} dr \, 
\frac{1}{(\nu^2 +3)(r-r_{-})(r-r_{+})} \Biggl[\le 2 \nu r - \sqrt{r_{+}r_{-}(\nu^2 +3)}  \ri \Biggr.
\nl
&&
\Biggl.
- 2 E \sqrt{\frac{\le 2 \nu r - \sqrt{r_{+}r_{-}(\nu^2 +3)}  \ri^2 -
  (\nu^2 +3)(r-r_{-})(r-r_{+})}{4E^2 +  (\nu^2 +3)(r-r_{-})(r-r_{+})}} \,\,\, 
  \Biggr]
   \, .
\eea
As in the non-rotating case, $t=0$ at $r=r_{\rm min}$, and so:
\beq
u (r_{\rm max}) - u(r_{\rm min}) = t_R + r^* (r_{\rm max}) - r^* (r_{\rm min}) \, .
\eeq
By direct computation, we find the relation
\bea
\frac{V}{4 \pi l^2}  & = & \int_{r_{\rm min}}^{r_{\rm max}} dr \, \left[ \frac{\sqrt{4E^2 +  (\nu^2 +3)(r-r_{-})(r-r_{+})}}
{2 (\nu^2 +3)(r-r_{-})(r-r_{+})}    \right.
\nl
& & \sqrt{ \le 2 \nu r - \sqrt{r_{+}r_{-}(\nu^2 +3)}  \ri^2 -  (\nu^2 +3)(r-r_{-})(r-r_{+})}
\nl & &
\left. - E \frac{2\nu r - \sqrt{r_{+}r_{-} (\nu^2 +3)}}{(\nu^2 +3)(r-r_{-})(r-r_{+})} \right] + E (u (r_{\rm max})- u(r_{\rm min})) \, .
\eea
Using the previous definitions and simplifying the expression, we obtain again the  result
\beq
\frac{dV}{d \tau} = 2 \pi l E \, ,
\label{rateV}
\eeq
where $\tau=l \, t_b=2 l \, t_R$. At large $\tau$, $E$ approaches the constant 
\beq
E_0= \frac{(r_{+}-r_{-})}{4} \sqrt{\nu^2 +3} \, .
\eeq
Numerical calculation are shown in figure  \ref{figura2}.
As a consistency check, putting $ \nu=1 $ for the BTZ case, we find
\beq
\lim_{\tau \rightarrow \infty} \frac{dV}{d \tau} = \pi l (r_+ - r_-) \, ,
\eeq
which is the same result found in standard coordinates on the Poincar\'e
 patch when we perform the change of variables (\ref{change-BTZ}).

The late time limit of the maximal volume slices
 can be found also in a simpler way, as in \cite{Stanford:2014jda}.
In this limit, we expect that the maximal volume slice sits at constant $r$,
due to translation invariance in time. We can then consider volume
slices at a constant $r=\hat{r}$. Extremizing the volume from the metric in eq. (\ref{BHole}),
we find that the only possible maximal constant-$r$ slice sits at
\beq
\hat{r}= \frac{r_{+}+r_{-}}{2} \, .
\eeq
Inserting this value back in the volume functional, we recover eq. (\ref{rateV})
with $E=E_0$.

\section{Conclusions}

\label{Conclu}

The result of our calculation gives that  the volume of the extremal slices in WAdS
is a monotonically growing function for $\tau>0$,  whose  late time growth rate approaches to
\beq
\frac{d V}{d \tau} \rightarrow \frac{\pi l}{2}
(r_+ -r_-)  \sqrt{3+\nu^2}  =S T \frac{8 \pi G l }{\sqrt{3+\nu^2}} \, ,
\eeq
where $S$ is the Bekenstein-Hawking entropy and $T$ the Hawking temperature.
The late time  rate  vanishes for extremal black holes ($ r_{+}=r_{-}$)
and is proportional to $T S$.

In AdS$_D$, we have that the coefficient of proportionality between
 complexity and volume \cite{Susskind:2014moa} is usually taken as:
\beq
 C = (D-1) \frac{V}{G l} \, .
\eeq
The late-time rate of growth of the volume  is:
\beq
\lim_{\tau \rightarrow \infty} \frac{d V}{d \tau} =\frac{8 \pi  G l}{D-1} S T  \, .
\eeq
For comparison, in the case of flat spacetime BHs, 
\beq 
\lim_{\tau \rightarrow \infty} \frac{d V}{d \tau} \approx \frac{G r_h}{D-3}  ST \, ,
\eeq
where $r_h$ is the horizon radius ($\approx$ refer to a neglected order one prefactor \cite{Susskind:2014moa}).
Consequently,  the proportionality coefficient between the late time rate of growth
of the volume and $TS$ depends on the kind of asymptotic of the spacetime.

In order to compare with the AdS$_3$ case, we can write the rate of growth of the volume in WAdS as:
\beq
\frac{d V}{d \tau} \rightarrow 
S T \,  4 \pi G l \, \eta \,  , \qquad \eta= \frac{2}{\sqrt{3+\nu^2}} \, .
\eeq
We may interpret the details of this result in distinct ways, depending 
on the exact holographic dictionary that we may conjecture between volume and complexity.
For example, it could be that complexity approaches at 
 late time  to $\eta \, TS$ (note
that $\eta \leq 1$  if we impose $\nu^2 \geq 1$);
if this is true, warping would make complexity rate decreases.
On the other hand, it could also  be that in spaces with WAdS asymptotic
 the holographic dictionary between
complexity and volume is changed by some non-trivial function of the warping parameter $\nu$;
for example, if we would have that
\beq
C=  \frac{2 }{G l \eta} V  \, ,
\eeq
the asymptotic complexity increase rate would be still $TS$ for every $\nu$.
 It would be interesting to find arguments 
 in order to be able to discriminate among these various possibilities.

From general considerations based on limits about the speed of computation,
 there is a  conjectured bound on the growth rate of complexity of a physical system \cite{Lloyd}.
This is called Lloyd's bound and states that 
\beq
\frac{dC}{d\tau} \leq \frac{2 E}{\pi } \, ,
 \label{lloyd}
\eeq
where \emph{E} is the energy associated to the physical system and units $\hbar=1$ are used.
In the case of BHs, the energy is identified with the BH mass, $E=M$.
In general relativity the definition of mass $M$ depends on the choice
of the asymptotic Killing vector used to define the conserved quantity.
In the WAdS case,
the boundary spatial direction $\theta$ in the metric (\ref{BHole}) 
is selected by the $r^2$ divergence at $r \rightarrow \infty $ in the  $d \theta^2$ coefficient of the metric.
Our definition of mass $M$ in eq. (\ref{M guess})
is in term of the Killing vector $\xi_t=2 \frac{\p}{\p t}$;
this choice is natural, because the norm of $\xi_t$ does not diverge at the boundary.
If we would choose another Killing vector to define the mass
\beq
\xi_j=\xi_t - j \frac{\p}{\p \theta} \, ,
\eeq
then, for non-zero $j$,
the norm of $\xi_j$ would diverge as $r^2$ for $r \rightarrow \infty$.
Instead in the AdS case ($\nu=1$)  the $r^2$ divergence of the metric at infinity disappears
and $\xi_t$ is not a natural time Killing vector, because it is not an eigenvector of the boundary metric.

Since the late-time complexity rate is proportional to $TS$, if we want that an universal
Lloyd's bound holds, we must require that some positive constant $k$ exists such that
 $k M \geq TS$ for every value of $r_+$, $r_-$  at fixed $\nu$.
 If we impose this, we find that
$k \geq g(\nu)$, where 
\beq
g(\nu)=\frac{6 \nu  \sqrt{\nu ^2-1}}{2 \sqrt{3} \nu  \left(2 \nu -
n\right)
   +3 n \sqrt{\left(\nu ^2-1\right) }} \, , \qquad    n=\sqrt{ 7 \nu ^2+4  \sqrt{3} \sqrt{\nu ^2-1} \nu -3} \, ,
\eeq
which is a decreasing function  of $\nu$, with $g(1) \rightarrow \infty$
and $g(\infty)  \approx 1.15 $. So it is possible, with appropriate normalization,
to introduce a Lloyd's bound proportional to $M$ for $\nu>1$. 
For $\nu=1$ (the AdS case) instead this is not possible; but indeed we know
that in this case  $M$ is not the correct mass because $\frac{\p}{\p t}$
does not correspond to the natural time direction in AdS; in this case 
there is a conjectured Lloyd's bound in terms of the usual time direction in AdS \cite{Brown:2015lvg}.
For $\nu<1$ we do not expect a Lloyd's bound, because there are closed time-like curves in the geometry;
 in this range the mass $M$ can  even be negative.

Several problems are left for further investigation:
\begin{itemize}
\item In this note we considered the case in which WAdS BHs are realized as solutions of
Einstein gravity with some appropriate matter content.  These objects can also be realized as
vacuum solution of TMG and NMG. In these cases  we expect some higher order corrections to the  CV
conjecture, analog to the area corrections in the Wald entropy formula. Some proposals have been discussed in
\cite{Alishahiha:2015rta,Alishahiha:2017hwg}. It would be interesting to compute these corrections explicitly.
\item The CA conjecture should also be investigated  for the WAdS BH solutions,
 both in the asymptotic rate of growth and in the initial transient period.
This was initiated in \cite{Ghodrati:2017roz} for the case of TMG; in this case the late-time complexity rate
is not proportional to $T S$, but it still vanishes in the extremal case.
\item It would be interesting to study complexity  in the boundary 
 Warped CFTs.  Lagrangian examples of free Warped CFT were
  introduced in \cite{Hofman:2014loa,Jensen:2017tnb}. 
\end{itemize}

 \section*{Acknowledgments}

We are grateful  to Shira Chapman for  useful comments on the manuscript.


\section*{Appendix}
\addtocontents{toc}{\protect\setcounter{tocdepth}{1}}
\appendix

\section{An explicit model}
\label{explicit-model}

In this appendix we consider an explicit model whose entropy
 satisfies the area law and admitting the metric eq.~(\ref{BHole}) as a solution \cite{Banados:2005da}.
This is a model of Einstein gravity in (2+1) dimensions minimally coupled to a gauge field with Chern-Simons and Maxwell terms:
\beq
S=  \frac{1}{16 \pi G}  \int d^3 x \left\{\sqrt{g} \left[ 
\left( R + \frac{2}{L^2}\right) -\frac{\kappa}{4} F^{\mu \nu} F_{\mu \nu}
\right] -\frac{\a}{2} \epsilon^{\mu \nu \rho} A_\mu F_{\nu \rho}\right\} \, ,
\eeq
where $\epsilon^{\mu \nu \rho} $ is the Levi-Civita tensorial density.
Here we put a coefficient $\kappa=\pm 1$ in front of  the Maxwell kinetic term.  
The equations of motion for the gauge field are
\beq
D_\mu F^{\a \mu} = -\frac{\a}{\kappa}  \frac{ \epsilon^{\a \nu \rho} }{\sqrt{g}} F_{\nu \rho} \, ,
\eeq
while the Einstein equations are
\beq
G_{\mu \nu} - \frac{1}{L^2} g_{\mu \nu} = \frac{\kappa}{2}  T_{\mu \nu} \, , \qquad
T_{\mu \nu} =   F_{\mu \a} F_{\nu}^{\,\,\, \a}-\frac14 g_{\mu \nu}
F^{\a \b} F_{\a \b} \, .
\eeq
We consider the set of coordinates $ (r,t, \theta) $ where the metric assumes the form (\ref{BHole}), and we choose a 
 gauge motivated by the ansatz from \cite{Banados:2005da}:
\beq
A= a dt+ (b+c r) d \theta \, , \qquad F = c \, dr \wedge d \theta \, ,
\label{ansatz per campo di gauge}
\eeq
where $ \lbrace a,b,c \rbrace $ is a set of constants.

In this gauge, the Maxwell equations give:
\beq
\a = \kappa \frac{\nu}{l} \, .
\eeq
From the Einstein equations,
in order to require absence of closed time-like curves ($\nu>1$), 
we have to choose $\kappa=-1$ and the following value of the parameters:
\beq
L = l \sqrt{\frac{2}{3-\nu^2} } \, , \qquad
c =l \sqrt{\frac{3}{2 } (\nu^2-1)}
\, , \qquad
\eeq
which are simultaneously defined only when $1<\nu^2<3$.
So there is conflict between absence of closed time-like  curves
and presence of ghosts ($\kappa=-1$).

In ref. \cite{Banados:2005da} the conserved charges associated to asympthotic isometries of the black hole have been computed starting from the following form of the metric in the coordinates $ (\tilde{t}, \tilde{r}, \tilde{\theta}) $: 
\beq
ds^2 = p d\tilde{t}^2 + \frac{d\tilde{r}^2}{h^2-pq} + 2 h d\tilde{t} d\tilde{\theta} + q d \tilde{\theta}^2 \, ,
\label{forma metrica Godel}
\eeq
with functions given by
\beq
p(\tilde{r})= 8 G \mu  \, , \qquad 
q (\tilde{r}) = -\frac{4G \mathcal{J} }{\a} +2 \tilde{r} - 2 \frac{\gamma^2}{L^2} \tilde{r}^2 \, , \qquad
h (\tilde{r}) = -2 {\a} \tilde{r}  \, ,
\eeq
and $U(1)$ gauge field 
\beq
A=A_{\tilde{t}} d \tilde{t}+ A_{\tilde{\theta}} d \tilde{\theta} \, , \qquad
A_{\tilde{t}} (\tilde{r}) = \frac{{\a} ^2 L^2 -1}{\gamma {\a} L} + \zeta  \, , \qquad
A_{\tilde{\theta}} (\tilde{r})= - \frac{4G}{ {\a} } Q + \frac{2\gamma}{L} \tilde{r} \, ,
\eeq
where
\beq
\gamma = \sqrt{\frac{1-{\alpha}^2 L^2}{8 G \mu}} \, ,
\eeq
and $\zeta$ is a gauge constant.

We can put the metric (\ref{BHole}) in the form (\ref{forma metrica Godel}) by means of the coordinate change:
\beq
\tilde{t}  = \sqrt{\frac{l^3}{\omega}} t  \, , \qquad
\tilde{r}=r - \frac{\sqrt{r_+ r_- (\nu^2 +3)}}{2 \nu} \, , \qquad
\tilde{\theta} = \frac{\sqrt{\omega l^3}}{2} \theta \, ,
\eeq
where
\beq
\omega = \frac{\nu^2 +3}{2 \nu l} \le \nu (r_+ + r_-) - \sqrt{r_+ r_- (\nu^2 +3)} \ri \, .
\eeq
The previous set of transformations is such that the gauge field in the coordinates $ (t,r,\theta) $ can be written as $  A= a dt + (b+cr) d \theta , $ motivating the ansatz (\ref{ansatz per campo di gauge}). 

The quantities $ \mu, \mathcal{J}, Q $ appearing in the previous solution are respectively 
identified with the mass, angular momentum and charge of the black hole.
The equations of motion and the change of coordinates do not uniquely fix the charge \emph{Q}, while we identify
\beq
\mu= \frac{\nu^2 +3}{16 G l^2} \le r_+ + r_- - \frac{\sqrt{r_+ r_- (\nu^2 +3)}}{\nu} \ri \, ,
\eeq
\beq
\mathcal{J} =  \frac{ 2 \nu  (r_+ + r_-) \sqrt{r_+ r_- (\nu^2 +3)} - (5\nu^2 +3)r_+ r_-}  
{8 G l  \le \nu  (r_+ + r_-)  - \sqrt{r_+ r_- (\nu^2 +3)}  \ri} \, .
\eeq
As it is pointed out in \cite{Banados:2005da}, the set $ \lbrace \mu, \mathcal{J} , Q \rbrace $ satisfies the first law of thermodynamics in the form
\beq
d \mu = T dS + \Omega d \mathcal{J} + \Phi_{\mathrm{tot}} dQ \, ,
\eeq
where the total electric potential is shown to be $ \Phi_{\mathrm{tot}} = 0, $ thus eliminating the contribution from the charge of the black hole.

This special form of the first law of thermodynamics is a consequence of the choice of the Killing vectors associated to mass and angular momentum in \cite{Banados:2005da}, since all the contributions coming from the charge are eliminated.

A direct match with the mass \emph{M} and angular momentum \emph{J} coming from the thermodynamic analysis in $ (t,r,\theta) $ coordinates gives:
\beq
\mu=\frac{M}{l^2} \, , \qquad \mathcal{J}=-\frac{4 J}{ \omega l^2} \, .
\eeq
In order to get the conserved charges associated to isometries in $ (t, r, \theta) $ coordinates, we need to adjust the normalization conditions:
\begin{itemize}
\item
The angular range $0 \leq \theta \leq 2 \pi $
corresponds to $0 \leq \tilde{\theta} \leq 2 \pi \frac{\sqrt{\omega l^3}}{2} $,
so extensive quantities, such as  mass, entropy and angular momentum
 in $ (t, r, \theta) $ coordinates get an extra  $\frac{\sqrt{\omega l^3}}{2}$ factor if we want to preserve the length of the integration along $ [0, 2 \pi] . $
\item Killing vectors are transformed as:
\beq
\frac{\p}{\p t} =\sqrt{\frac{l^3}{\omega}} \frac{\p}{\p \tilde{t} }  \, , \qquad
\frac{\p}{\p \theta} =\frac{\sqrt{\omega l^3}}{2 } \frac{\p}{\p \tilde{\theta} }  \, .
\eeq
\item In \cite{Banados:2005da} it is defined $ \Omega = - h(r_{+})/q(r_+) , $ 
while in eq. (\ref{M guess}),(\ref{J guess}) we followed the conventions of \cite{Anninos:2008fx}, where an additional factor of \emph{l} is put in the denominator both for the angular velocity and the Hawking temperature.
Choosing the last normalization amounts to modify $ \mu \rightarrow \mu/l , $ with the other conserved charges of the black hole unchanged.
\end{itemize}
Taking into account all these corrections, we get that the mass in $ (t,r, \theta) $ coordinates with the  Killing $\frac{\p}{\p t}$  is $M/2$ and the angular momentum  associated to the Killing $ -\frac{\p}{\p \theta}$ is   $J$. 
The $1/2$ factor in the normalization of the mass is reminiscent of Komar's anomalous factor and it is also pointed out for similar computations in \cite{Bouchareb:2007yx}.

\end{document}